\documentclass[12pt]{iopart}

\usepackage{float}
\usepackage{textcomp}
\usepackage{graphicx}
\usepackage{subfigure}
\usepackage{setspace}

\begin{document}

\title{Influence of kinetic energy on the metrology of Rabi frequency}

\author{Xingyu Zhang$^1$,
Xiaoguang Wang$^1$}
\address{$^1$Zhejiang Institute of Modern Physics, Department of Physics, Zhejiang University, Hangzhou, Zhejiang 310027, China}

\ead{xgwang1208@zju.edu.cn}
\vspace{10pt}

\begin{abstract}
The interacting model describing the Rabi transition is essential in studying atom-photon interactions, where the kinetic energy term is often neglected for the convenience of analysis. We first study the approximation through the fidelity approach and verify its valid region in the parameter
space of detuning and momentum. We find that as the radiation field's momentum and the absolute value of detuning decrease, the approximation becomes
valid. We further discover that the omission of the kinetic energy term will overestimate the measuring accuracy of the Rabi frequency in some parameter regions and underestimate the precision in other regimes with Fisher information tools' help. Notably, a specific choice of the initial Gaussian state's variance in position space will improve the measuring accuracy when we take the kinetic energy term into account. We also study the realistic measurement of the Rabi frequency for cases with and without the kinetic energy.
\end{abstract}

%
%
%
%
%
%

\section{Introduction}\label{section1}
J-C model, describing the interaction between light and matter \cite{1,2,3}, has been studied for many years.
However, in the original J-C model, the homogeneity of the optical fields has always been assumed.
As a result, the optical field is often thought to be independent of the position and evaluated
at the atomic center-of-mass location. So in this model, the motional state of the atom,
e.g., the center-of-mass motion, has been omitted directly. Nevertheless, since many systems
deal with both internal energy levels and the external motional states
\cite{2,4,5,6,7}, we wonder if the kinetic energy term influences the
measurement of the Rabi frequency?

Quantum Fisher information (QFI), $F$, corresponding to the classical
Fisher information (CFI) which is widely used in statistics and other
fields \cite{8,9}, is the key indicator of how useful a quantum
state is for quantum metrology. It can give a lower bound on the sensitivity
$\Delta^{2}\phi\geq1/F$ for a parameter $\phi$ \cite{10,11,12,13}.
So if we can get larger quantum Fisher information, we will gain a
more precise estimation of the parameter. That is the spirit of
quantum metrology, and the calculation of QFI has been studied for
many years \cite{14,15,16,17}. We can see the future application
of the Fisher information in quantum sensing \cite{18,19,20}, too.

Considering the importance of the Rabi frequency \cite{21,22,35,36,37},
precise measurement of the Rabi frequency is crucial, which is
the main discussion in this article. From the perspective of QFI, we try to understand
the influence of the kinetic energy term on measuring Rabi frequency under
different parameters. In numerous studies, we have noticed that the kinetic
energy term is omitted in some cases \cite{6}
(e.g., Raman-Nath regime). However, for many cases, it can not be omitted directly
(Bragg regime). We try to figure out the mechanism in a general light-atom
interaction Hamiltonian.

Fisher  information has been used in many atom-light interaction cases, including fundamental two-level
system dynamics \cite{23,24,25}, Raman transition \cite{26,27}, or the spin squeezing context \cite{28}. However, either they did not consider
the metrology on Rabi frequency, or the effect of motional state was excluded.
Here we use the Hamiltonian considering the inhomogeneous field,
and this means that the spatial motion of the atomic center of mass
plays an important role in the interacting Hamiltonian between the
field and the atom \cite{29,30,31,32,33,34}. We deal with this kind of Hamiltonian with or without
the kinetic energy term to figure out whether this term is crucial
for the metrology of Rabi frequency. We have found that in some regimes, e.g., the detuning is
larger than zero at a given wave vector, ignoring the kinetic term will increase the QFI,  which means we can measure the
Rabi frequency more precisely, but the neglect will decrease the QFI
in other areas. In addition, we work out the
corresponding CFIs to see how the specific measurements match the
theoretical maximum. Our analytical results will help determine when
the omission is valid in different regimes of parameters, and the comparison
between QFI and CFI will make us know if some specific measurements can
attain the best sensitivity under certain parameters.

The main content of this article is divided into four parts. The first part is
Section \ref{sec:2}, where we compare the final states with or without kinetic
energy term directly through fidelity. The above two cases become
similar when the momentum of radiation field and the absolute value of the detuning
go down, which means the kinetic energy becomes less critical. Then,
in Section \ref{sec:3}, we give our main results about the influence
on QFI when the kinetic energy term is eliminated. We compare the
two cases with or without the kinetic energy term via changing the
parameters. The Section \ref{sec:4} mainly talks about the possibility of attaining
QFI in measurement (CFI) without the kinetic energy term. The Section \ref{sec:5} discusses QFI and CFI when the kinetic energy term is present
\cite{38,39}.

Actually, we mainly choose the detuning and the wave vector to see
the effect of kinetic energy term. It can be seen that the decrease
in both two parameters can increase the QFI and CFI. However, there
are also other parameters that influence the result. It is evident
that the bigger the atomic mass is, the less critical the kinetic
energy term is. Besides, if the initial state is Gaussian type in
the position space, the appropriate choice of the variance of the
Gaussian state will enhance the Fisher information while taking the
kinetic energy into account.

\section{Direct comparison of final states with respect to kinetic energy
term}\label{sec:2}
In this section, we will briefly discuss the effect of kinetic energy
term in Hamiltonian on the fidelity. We will utilize the two-level system
interacting with light and set $|b\rangle$ and $|a\rangle$ as the upper and lower states, respectively. We set the lower energy of $|a\rangle$ as zero and the general form of Hamiltonian under the rotating wave approximation is
\begin{equation}
\hat{H}=\frac{\hat{p}^{2}}{2m}+\hbar\omega_{ba}|b\rangle\langle b|+\frac{\hbar\Omega}{2}(|b\rangle\langle a|\mathrm{e}^{\mathrm{i}(k_{0}\hat{z}-\omega_{L} t)}+|a\rangle\langle b|\mathrm{e}^{-\mathrm{i}(k_{0}\hat{z}-\omega_{L} t)}),
\end{equation}
where the $\omega_{ba}$ is the level difference, $k_0$ is the wave vector, and $\omega_{L}$ is the laser frequency.
The first two terms are the motional and internal Hamiltonian of the atom, respectively, and the last term represents the interaction between the running wave and the atom. Such kind of Hamiltonian is the semi-classical version of that in \cite{34}, and can be also found in \cite{2,38}. In the frame rotating at the laser frequency, i.e., using the unitary transformation $\hat{U}=\mathrm{e}^{-\mathrm{i}\omega_{L}t|b\rangle\langle b|}$, our Hamiltonian will be written as
\begin{equation}
\hat{H}_{1}=\frac{\hat{p}^{2}}{2m}+\hbar\Delta|b\rangle\langle b|+\frac{\hbar\Omega}{2}(|b\rangle\langle a|\mathrm{e}^{\mathrm{i}k_{0}\hat{z}}+|a\rangle\langle b|\mathrm{e}^{-\mathrm{i}k_{0}\hat{z}}),\label{eq:Hwithkinetic}
\end{equation}
where $\Delta=\omega_{ba}-\omega_{L}$. The parameters are initially set as $\hbar\sim10^{-34}\ \mathrm{J\cdot s},\ \Omega\sim10^{3}\ \mathrm{s^{-1}},\ m\sim1.44\times10^{-25}\ \mathrm{kg}$ \cite{38,39,40,41}, and we will constantly use these parameters below. We will see below these initial parameters are not good enough for the metrology of the Rabi frequency. However, we will figure out the better region of specific parameters to enhance the measuring precision.

In order to simplify the calculations under the evolution of the above
Hamiltonian, we need to use a unitary transformation \cite{34}
\begin{equation}
\hat{W}=|b\rangle\langle b|\mathrm{e}^{\mathrm{i}\frac{k_{0}\hat{z}}{2}}+|a\rangle\langle a|\mathrm{e}^{-\mathrm{i}\frac{k_{0}\hat{z}}{2}}
\end{equation}
to get a $su$(2)-type effective Hamiltonian

\begin{eqnarray}
\hat{H}_{e1} & =\hat{W}^{\dagger}\hat{H}_{1}\hat{W}\nonumber \\
 & =\frac{\hat{p}^{2}}{2m}+\frac{\hbar^{2}k_{0}^{2}}{8m}+\frac{\hbar\Delta}{2}+(\frac{\hbar k_{0}\hat{p}}{2m}+\frac{\hbar\Delta}{2})\sigma_{z}+\frac{\hbar\Omega}{2}\sigma_{x}.
\end{eqnarray}
Then the evolution operator can be written as
\begin{equation}
\hat{U}_{1}=\mathrm{e}^{-\mathrm{i}\frac{t}{\hbar}\hat{H}_{1}}=\hat{W}\mathrm{e}^{-\mathrm{i}\frac{t}{\hbar}\hat{H}_{e1}}\hat{W}^{\dagger}.
\end{equation}

The advantage of the above effective Hamiltonian is that it has eliminated
the $\hat{z}$ spatial component, which does not commute with $\hat{z}$-direction
momentum $\hat{p}$. We then try to omit the kinetic energy term. This situation is suitable for relatively large coupling strength, big atomic mass, or shorter time evolution. The Hamiltonian can be written as
\begin{equation}
\hat{H}_{2}=\hbar\Delta|b\rangle\langle b|+\frac{\hbar\Omega}{2}(|b\rangle\langle a|\mathrm{e}^{\mathrm{i}k_{0}\hat{z}}+|a\rangle\langle b|\mathrm{e}^{-\mathrm{i}k_{0}\hat{z}}),\label{eq:Hwithout}
\end{equation}
After the same unitary transformation
\begin{equation}
\hat{W}=|b\rangle\langle b|\mathrm{e}^{\mathrm{i}\frac{k_{0}\hat{z}}{2}}+|a\rangle\langle a|\mathrm{e}^{-\mathrm{i}\frac{k_{0}\hat{z}}{2}},
\end{equation}
we get an effective $su$(2)-type Hamiltonian

\begin{eqnarray}
\hat{H}_{e2} & =\hat{W}^{\dagger}\hat{H}_{2}\hat{W}\nonumber \\
 & =\frac{\hbar\Delta}{2}+\frac{\hbar\Delta}{2}\sigma_{z}+\frac{\hbar\Omega}{2}\sigma_{x}
\end{eqnarray}

\begin{figure}[H]
\centering \includegraphics[width=12cm,height=8cm]{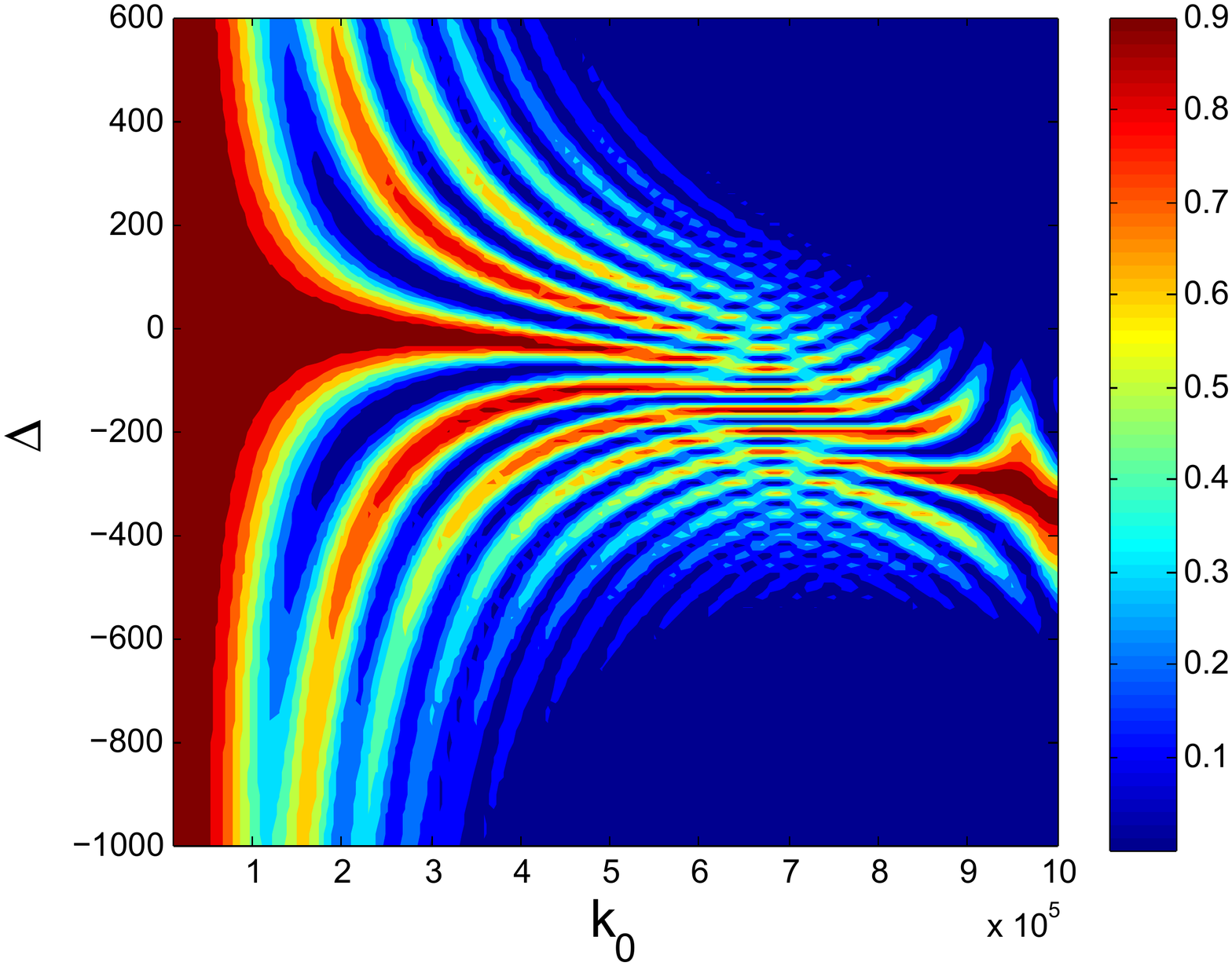}
\caption{Contour plot of the fidelity between final states of the two cases.}
\label{Fig1}
\end{figure}

This effective Hamiltonian does not even consist of the momentum operator, which means we can obtain a more straightforward expression. Then we have the evolution operator
\begin{equation}
\hat{U}_{2}=\mathrm{e}^{-\mathrm{i}\frac{t}{\hbar}\hat{H}_{2}}=\hat{W}\mathrm{e}^{-\mathrm{i}\frac{t}{\hbar}\hat{H}_{e2}}\hat{W}^{\dagger}.
\end{equation}
Before calculating the QFI, we first
compare the two final states under the evolution of the two Hamiltonians. Here, we assume the initial state is
$|\psi_{in}\rangle=|a\rangle|\psi_{0}\rangle$, where the motional
state in position space is
\begin{equation}
\langle z|\psi_{0}\rangle=\exp(-z^{2}/2\sigma^{2})/(\pi\sigma^{2})^{1/4},\label{eq:initial}
\end{equation}
where $\sigma\sim4\times10^{-5}\ \mathrm{m}$ \cite{38}.
To show the exact influence of the kinetic energy term on the evolved final
state, we calculate the fidelity
\begin{eqnarray}
\mathcal{F} & =|\langle\psi_{f1}|\psi_{f2}\rangle|^{2}\nonumber \\
 & =|\langle\psi_{in}|\hat{U}_{1}^{\dagger}\hat{U}_{2}|\psi_{in}\rangle|^{2}\nonumber \\
 & =|\langle\psi_{0}|\langle a|\hat{W}\mathrm{e}^{\mathrm{i}\frac{t}{\hbar}\hat{H}_{e1}}\mathrm{e}^{-\mathrm{i}\frac{t}{\hbar}\hat{H}_{e2}}\hat{W}^{\dagger}|a\rangle|\psi_{0}\rangle|^{2},
 \label{eq:fidelity}
\end{eqnarray}
and give the contour plot in figure \ref{Fig1} (refer to \ref{appendixB} for details). We can see the
general trend: the fidelity becomes larger as the absolute
value of the $\Delta$ or $k_{0}$ goes down, leading to the valid omission of kinetic energy
in the Hamiltonian. We will see below this trend basically satisfies our calculation of QFI.

\section{The effect of kinetic energy term on QFI}\label{sec:3}
Now let us turn back to the calculation of the QFI. For the $su$(2)-type
Hamiltonian, a general solution for the QFI has been discussed \cite{14}.
We can define a Hermitian operator independent of the initial state
\begin{equation}
\mathcal{H}_{1}=\mathrm{i}(\partial_{\Omega}\hat{U}_{1}^{\dagger})\hat{U}_{1}.
\end{equation}
If the initial state is pure state, the QFI can be expressed as
\begin{equation}
F=4\langle\psi_{in}|\Delta^{2}\mathcal{H}_{1}|\psi_{in}\rangle.\label{eq:quantfisher}
\end{equation}
In fact, the momentum representation will facilitate our computation,
as will see in \ref{appendixA}. We still use the initial state (\ref{eq:initial}).
After the evolution of the Hamiltonian (\ref{eq:Hwithkinetic}),
we finally get the integral formalism of QFI
\begin{eqnarray}
F & =4\Biggl[\frac{\sigma}{\hbar\sqrt{\pi}}\int_{-\infty}^{\infty}\mathrm{d}p\ \mathrm{e}^{-\frac{p^{2}\sigma^{2}}{\hbar^{2}}}\Biggl\{\left(\frac{\Omega^{2}\left(\sin\left(\omega t\right)-\omega t\right)}{2\omega^{3}}-\frac{\sin\left(\omega t\right)}{2\omega}\right)^{2}\nonumber \\
 & +\left(\frac{\sqrt{\omega^2-\Omega^2}\left(1-\cos\left(\omega t\right)\right)}{2\omega^{2}}\right)^{2}+\left(\frac{\Omega\sqrt{\omega^2-\Omega^2}(\sin\left(\omega t\right)-\omega t)}{2\omega^{3}}\right)^{2}\Biggl\}\Biggl]\nonumber \\
 & -4\Biggl[\frac{\sigma}{\hbar\sqrt{\pi}}\int_{-\infty}^{\infty}\mathrm{d}p\ \mathrm{e}^{-\frac{p^{2}\sigma^{2}}{\hbar^{2}}}\Biggl(\frac{\Omega\sqrt{\omega^2-\Omega^2}(\sin\left(\omega t\right)-\omega t)}{2\omega^{3}}\Biggr)\Biggr]^{2},
\end{eqnarray}
where
\begin{equation}
\omega=\sqrt{\left(\Delta+\frac{k_{0}\left(\frac{k_{0}\hbar}{2}+p\right)}{m}\right)^{2}+\Omega^{2}}.
\end{equation}

Again, as for the case without the kinetic energy term, we can
define a Hermitian operator independent of the initial state
\begin{equation}
\mathcal{H}_{2}=\mathrm{i}(\partial_{\Omega}\hat{U}_{2}^{\dagger})\hat{U}_{2}.
\end{equation}

The whole procedure is similar to the above case, and we will continue to use the
initial state $|\psi_{in}\rangle=|a\rangle|\psi_{0}\rangle$. Notably, when the kinetic
energy term is absent, the calculation will be much easier, and we
can get an analytical expression for the QFI written in \ref{appendixA}.
After the evolution of the Hamiltonian (\ref{eq:Hwithout}),
the QFI at $t$ will be:
\begin{eqnarray}
F=\frac{\left(\Omega^{2}\omega't+\Delta^{2}\sin\left(\omega't\right)\right)^{2}+\Delta^{2}\omega'^{2}\left(\cos\left(\omega't\right)-1\right)^{2}}{\omega'^{6}},
\end{eqnarray}
where
\begin{equation}
\omega'=\sqrt{\Delta^{2}+\Omega^{2}}.
\end{equation}

\begin{figure}[H]
\centering
\subfigure[]{\includegraphics[width=6cm,height=5cm]{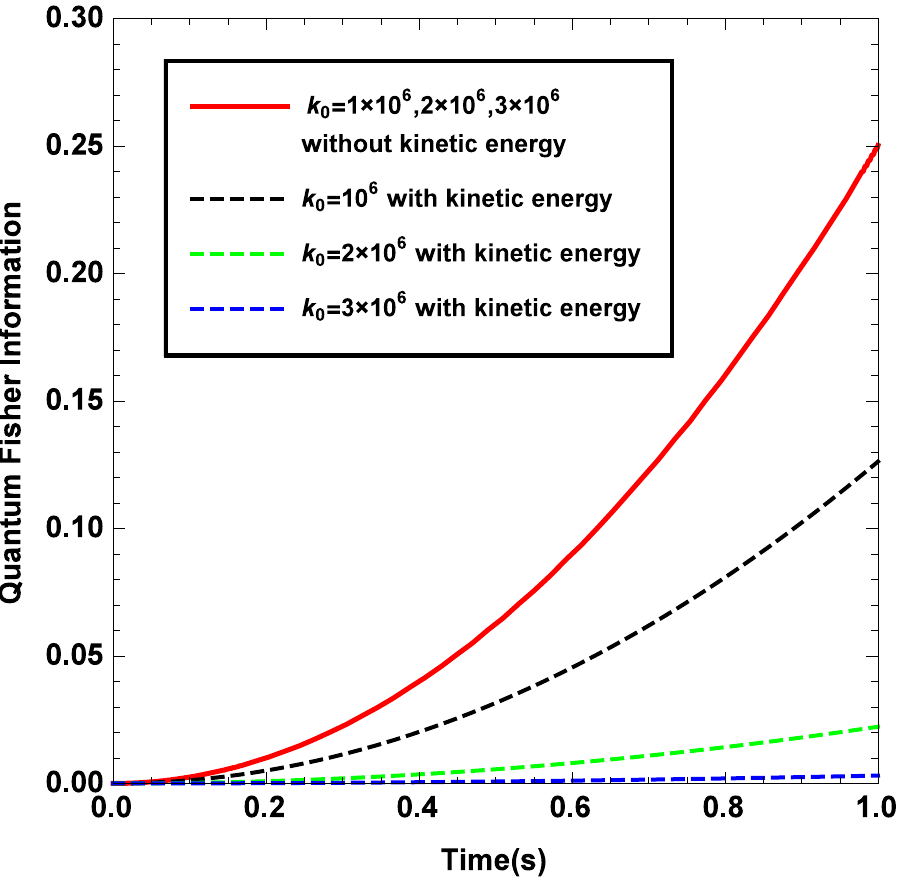}\label{Fig2a}}
\subfigure[]{\includegraphics[width=6cm,height=5cm]{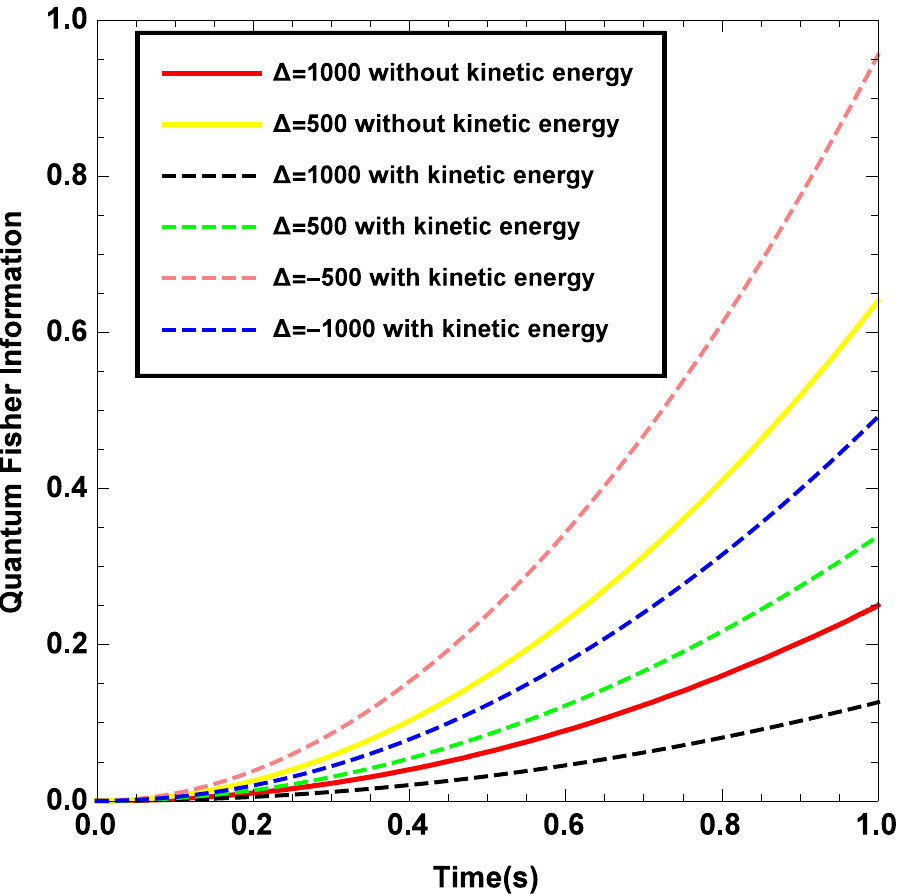}\label{Fig2b}}
\caption{(a) QFI with (dashed line) or without (solid lines) kinetic energy
using $k_{0}=10^{6},2\times10^{6},2\times10^{6}~\mathrm{m}^{-1}$
while setting $\Delta=1000\ \mathrm{s^{-1}}$. (b) QFI with (dashed
lines) or without (solid lines) kinetic energy using $\Delta=1000,500,100~\mathrm{s}^{-1}$
while setting $\ k_{0}=10^{6}\ \mathrm{m^{-1}}$.}
\label{Fig2}
\end{figure}

We compare it with the QFI with kinetic energy term in figure \ref{Fig2} concerning the detuning and light's momentum. Obviously, the QFI without the kinetic energy term does not dependent on $k_{0}$. When the kinetic energy term is taken into account, enhancing $k_{0}$ will lower the QFI drastically. We also conclude from the figures that if the detuning is positive at the given wave vector, the QFI without the kinetic energy term is always larger
than that with the kinetic energy term, which means we will overestimate the precision of the Rabi frequency if we omit the kinetic energy. However, if the detuning is negative, the situation will become different and we will discuss it later. In addition, we can see that decreasing the absolute value of the detuning $\Delta$ will increase the QFI in both cases.

\begin{figure}[H]
\centering \includegraphics[width=7cm,height=4cm]{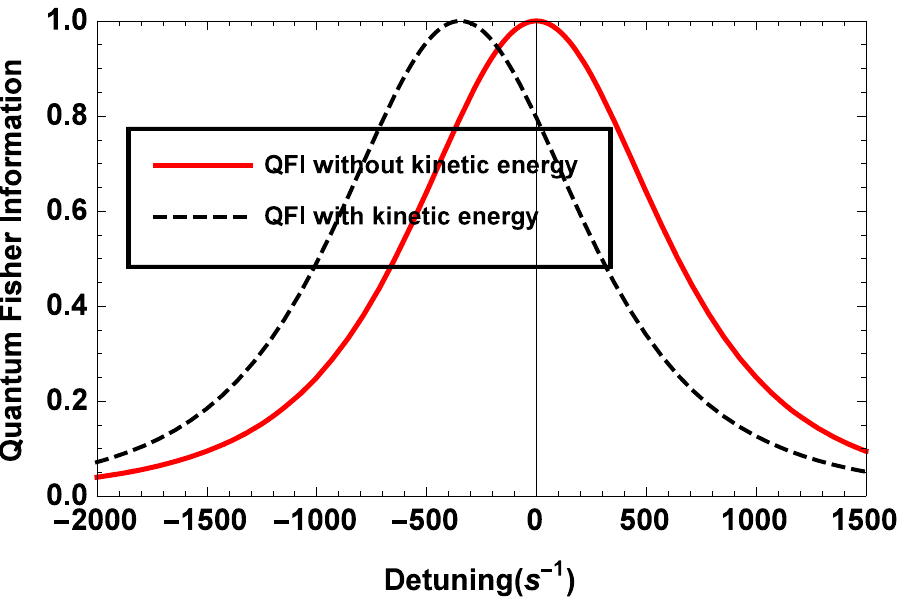}
\caption{QFI with respect to $\Delta$, here we set $t=1\mathrm{s}$ and$\ k_{0}=10^{6}\ \mathrm{m^{-1}}$.}
\label{Fig3}
\end{figure}

In order to see the influence of the detuning on the QFI more precisely, we plot the curve of QFI with respect to the detuning in figure \ref{Fig3}. We can obtain the fact that the QFI without the kinetic energy term depends only on the absolute value of the detuning while that with the kinetic energy has an axis of symmetry slightly deviating from zero. That is the reason why the QFI without the kinetic energy term is larger when $\Delta>0$. When the detuning is negative and smaller than a specific value (about -200$\mathrm{s}^{-1}$) at the given $k_{0}$, the QFI without kinetic energy term is smaller than that with kinetic energy. It means we will underestimate the measurement precision if we neglect the kinetic energy term when detecting the Rabi frequency.

\begin{figure}[H]
\centering \includegraphics[width=7cm,height=4cm]{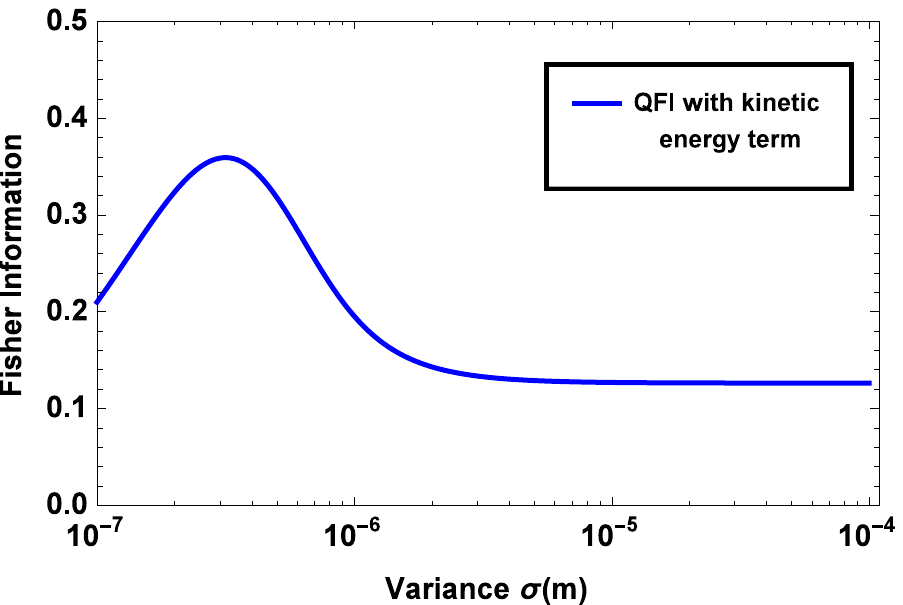}
\caption{QFI with respect to $\sigma$, here we set $t=1\mathrm{s}$ ,$\ k_{0}=10^{6}\ \mathrm{m^{-1}}$, and $\Delta=10^{3}\ \mathrm{s^{-1}}$.}
\label{Fig4}
\end{figure}

Nevertheless, unlike the case without the kinetic energy, in figure \ref{Fig4} we
show how the Gaussian state's variance $\sigma$ influences the QFI
with the kinetic energy term.

\section{Practical measurement of Rabi frequency without kinetic
energy}\label{sec:4}

In general, we do not know if the QFI will be saturated in practical measurement.
Although the QFI gives the theoretical best sensitivity, we still
need to choose a particular measurement scheme to attain it. The CFI
gives the sensitivity of a parameter (in our case $\Omega$), whose information is hidden
in the probability distribution of an observable's outcome. Therefore, we
will find a realistic observable and examine the parameter sensitivity provided by measuring the given observable values. For simplicity, we will
firstly check out the case without kinetic energy term in this section.

The definition of CFI in terms of a parameter $\theta$ is
\begin{equation}
I=\int P(\lambda|\theta)(\frac{\partial\ln P(\lambda|\theta)}{\partial\theta})^{2}\mathrm{d}\lambda,
\end{equation}
where $P(\lambda|\theta)$ is the probability of obtaining $\lambda$
when we measure the observable $\hat{\Lambda}$ with the parameter being $\theta$.
For example, if we carry out the population-difference measurement (PDM)
$\hat{\Lambda}=\hat{S}_{z}$, then the CFI will be denoted as
\begin{equation}
I_{PDM}=\Sigma_{i=a,b}(\partial_{\Omega}P_{i})^{2}/P_{i}.
\end{equation}
In the case of continuous variables, such as momentum measurement (MM)
$\hat{\Lambda}=\hat{p}$, the analytical expression of the corresponding
CFI is given below
\begin{equation}
I_{MM}=\int(\partial_{\Omega}P(p))^{2}/P(p)\mathrm{d}p.
\end{equation}

To calculate the CFI with the $\hat{S}_{z}$, since we already know
\begin{equation}
I_{PDM}=\Sigma_{i=a,b}(\partial_{\Omega}P_{i})^{2}/P_{i}=(\partial_{\Omega}P_{a})^{2}/[P_{a}(1-P_{a})],
\end{equation}
we can use $P_{a}=\langle\psi_{a}|\psi_{a}\rangle=|\langle a|\psi_{out}\rangle|^{2}$
to represent the probability of finding the final state in the internal state $|a\rangle$.
Due to the result
\begin{eqnarray}
|\psi_{a}\rangle & =\langle a|\mathrm{e}^{-\mathrm{i}\frac{t}{\hbar}\hat{H}_{2}}|a\rangle|\psi_{0}\rangle=\langle a|W\mathrm{e}^{-\mathrm{i}\frac{t}{\hbar}\hat{H}_{e2}}W^{\dagger}|a\rangle|\psi_{0}\rangle\nonumber \\
 & =\mathrm{e}^{-\mathrm{i}\frac{k_{0}\hat{z}}{2}}\langle a|\mathrm{e}^{-\mathrm{i}\frac{t}{\hbar}\hat{H}_{e2}}\mathrm{e}^{\mathrm{i}\frac{k_{0}\hat{z}}{2}}|a\rangle|\psi_{0}\rangle\nonumber \\
 & =\langle a|\mathrm{e}^{-\mathrm{i}\frac{t}{\hbar}\hat{H}_{e2}}|a\rangle|\psi_{0}\rangle,
\end{eqnarray}
the probability is
\begin{equation}
P_{a}=|\langle a|\mathrm{e}^{-\mathrm{i}\frac{t}{\hbar}\hat{H}_{e2}}|a\rangle|^{2}=\cos^{2}c_{1}+n_{z1}^{2}\sin^{2}c_{1},
\end{equation}
where
\begin{eqnarray}
c_{1}  =-\frac{t}{2}\omega',
n_{z1} =\frac{\Delta}{\omega'}.
\end{eqnarray}

\begin{figure}[H]
\centering \includegraphics[width=7cm,height=4cm]{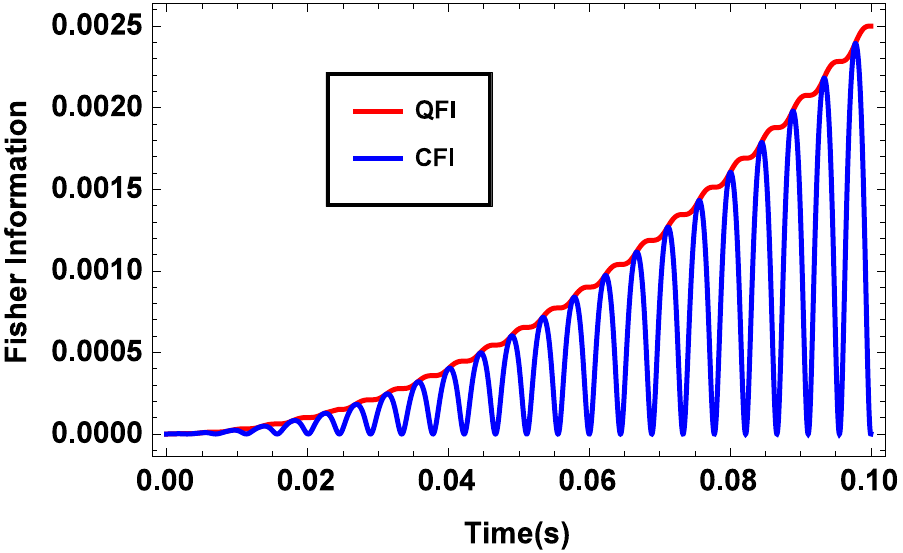}
\caption{Fisher Information, in which the parameters are defaultly set as$\ k_{0}=10^{6}\ \mathrm{m^{-1}}$,
and $\Delta=10^{3}\ \mathrm{s^{-1}}$.}
\label{Fig5}
\end{figure}

So we finally get our analytical solution of CFI:
\begin{equation}
I_{PDM}=\frac{2\left(2\Delta^{2}\omega'\sin\left(\frac{1}{2}\omega't\right)+\omega'^{2}t\Omega^{2}\cos\left(\frac{1}{2}\omega't\right)\right)^{2}}{\omega'^{6}\left(2\Delta^{2}+\Omega^{2}\cos\left(\omega't\right)+\Omega^{2}\right)}.
\end{equation}
In the following, we summarize the Fisher information (QFI, CFI for
$\hat{S}_{z}$ ) in figure \ref{Fig5}.

\begin{figure}[H]
\centering \subfigure[]{\includegraphics[width=7cm,height=4.5cm]{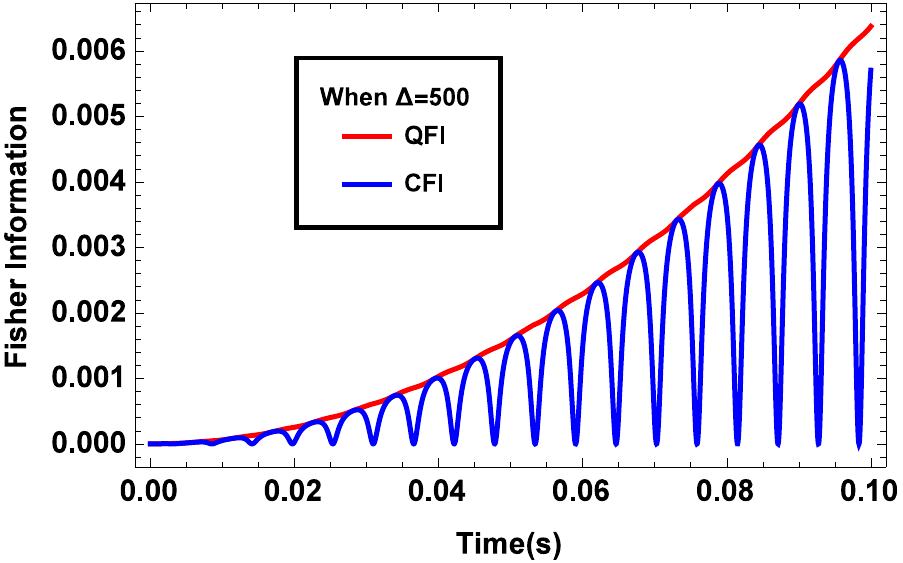}\label{Fig6a}}
 \subfigure[]{\includegraphics[width=7cm,height=4.5cm]{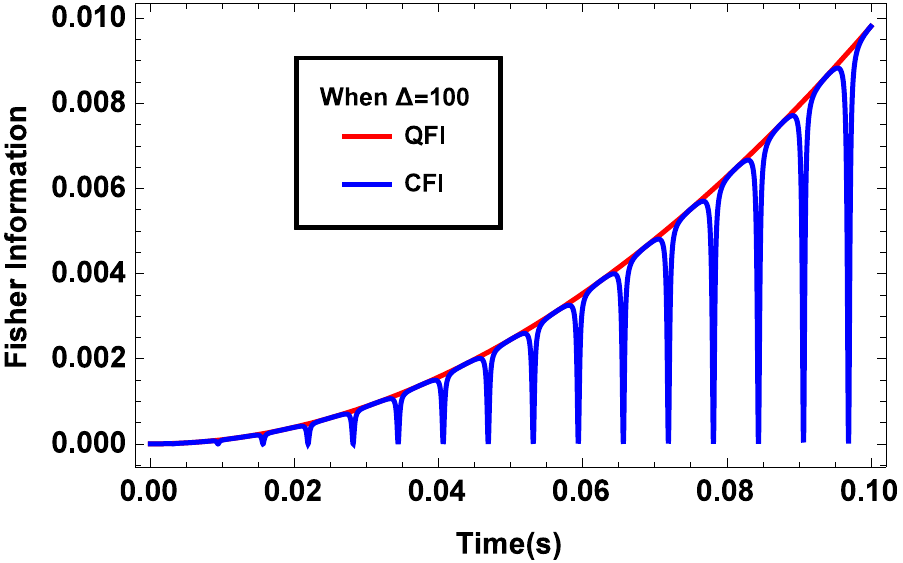}\label{Fig6b}}\\
 \caption{QFI and CFI when the kinetic energy term is absent for different detunings:
$\Delta=500,100~\mathrm{s^{-1}}$ while keeping$\ k_{0}=10^{6}\ \mathrm{m^{-1}}$. We can see that the Fisher information basically become
larger as the detuning decreases.}
\label{Fig6}
\end{figure}

We can see the Fisher information is so small that the final state
almost does not contain information of $\Omega$. The situations become
different when the parameters change, such as $\Delta$ for different
values in figure \ref{Fig6}.

We obtain when the detuning becomes sufficiently small, the Fisher
information becomes considerable. Fortunately, in the case where the
kinetic energy term is omitted, the QFI will be saturated only through
the population-difference measurement. Moreover, it appears that as the
detuning goes down, the CFI matches QFI with higher probability. Actually,
in the case of near resonance, the CFI of population-difference
measurement coincides precisely with the results of QFI.

\section{Practical measurement of Rabi frequency with kinetic energy}\label{sec:5}
We will now turn to the case with the kinetic energy term to see
if the QFI will be saturated. The same as the QFI, the CFI with the
kinetic energy term is a little bit cumbersome. To calculate $I_{PDM}=(\partial_{\Omega}P_{a})^{2}/[P_{a}(1-P_{a})]$,
we need to know $|\psi_{a}\rangle=\langle a|\psi_{out}\rangle$:
\begin{eqnarray}
|\psi_{a}\rangle &=\langle a|W\mathrm{e}^{-\mathrm{i}\frac{t}{\hbar}\hat{H}_{e1}}W^{\dagger}|a\rangle|\psi_{0}\rangle\nonumber \\
&=\langle a|\mathrm{e}^{-\mathrm{i}\frac{k_{0}}{2}\hat{z}}\mathrm{e}^{-\mathrm{i}\frac{t}{\hbar}\hat{H}_{e1}}\mathrm{e}^{\mathrm{i}\frac{k_{0}}{2}\hat{z}}|a\rangle|\psi_{0}\rangle\nonumber \\
&= \mathrm{e}^{-\mathrm{i}\frac{k_{0}}{2}\hat{z}}\mathrm{e}^{-\mathrm{i}\frac{t}{\hbar}(\frac{\hat{p}^{2}}{2m}+\frac{\hbar^{2}k_{0}^{2}}{8m}+\frac{\hbar\Delta}{2})}(\cos c_{2}-\mathrm{i}n_{z2}\sin c_{2})\mathrm{e}^{\mathrm{i}\frac{k_{0}}{2}\hat{z}}|\psi_{0}\rangle,
\end{eqnarray}
where the operators read
\begin{equation}
\hat{c}_{2}=-\frac{t}{\hbar}\sqrt{(\frac{\hbar\Omega}{2})^{2}+(\frac{\hbar k_{0}}{2m}\hat{p}+\frac{\hbar\Delta}{2})^{2}},
\end{equation}
\begin{equation}
\hat{n}_{z2}=\frac{\frac{\hbar k_{0}}{2m}\hat{p}+\frac{\hbar\Delta}{2}}{\sqrt{(\frac{\hbar\Omega}{2})^{2}+(\frac{\hbar k_{0}}{2m}\hat{p}+\frac{\hbar\Delta}{2})^{2}}}.
\end{equation}
As a result, we can get the probability:
\begin{eqnarray}
P_{a} & =\langle\psi_{0}|\mathrm{e}^{-\mathrm{i}\frac{k_{0}}{2}\hat{z}}(\cos^{2}c_{2}+n_{z2}^{2}\sin^{2}c_{2})\mathrm{e}^{\mathrm{i}\frac{k_{0}}{2}\hat{z}}|\psi_{0}\rangle\nonumber \\
 & =\int\mathrm{d}p\langle\psi_{0}|\mathrm{e}^{-\mathrm{i}\frac{k_{0}}{2}\hat{z}}(\cos^{2}c_{2}+n_{z2}^{2}\sin^{2}c_{2})\mathrm{e}^{\mathrm{i}\frac{k_{0}}{2}\hat{z}}|p\rangle\langle p|\psi_{0}\rangle\nonumber \\
 & =\int\mathrm{d}p|\langle\psi_{0}|p\rangle|^{2}(\cos^{2}c'+n_{z}'^{2}\sin^{2}c'),
\end{eqnarray}
where
\begin{equation}
c'=-\frac{t}{\hbar}\sqrt{(\frac{\hbar\Omega}{2})^{2}+(\frac{\hbar k_{0}}{2m}p+\frac{\hbar^{2}k_{0}^{2}}{4m}+\frac{\hbar\Delta}{2})^{2}},
\end{equation}
\begin{equation}
n_{z}'=\frac{\frac{\hbar k_{0}}{2m}p+\frac{\hbar^{2}k_{0}^{2}}{4m}+\frac{\hbar\Delta}{2}}{\sqrt{(\frac{\hbar\Omega}{2})^{2}+(\frac{\hbar k_{0}}{2m}p+\frac{\hbar^{2}k_{0}^{2}}{4m}+\frac{\hbar\Delta}{2})^{2}}}.
\end{equation}

\begin{figure}[H]
\centering \includegraphics[width=7cm,height=4cm]{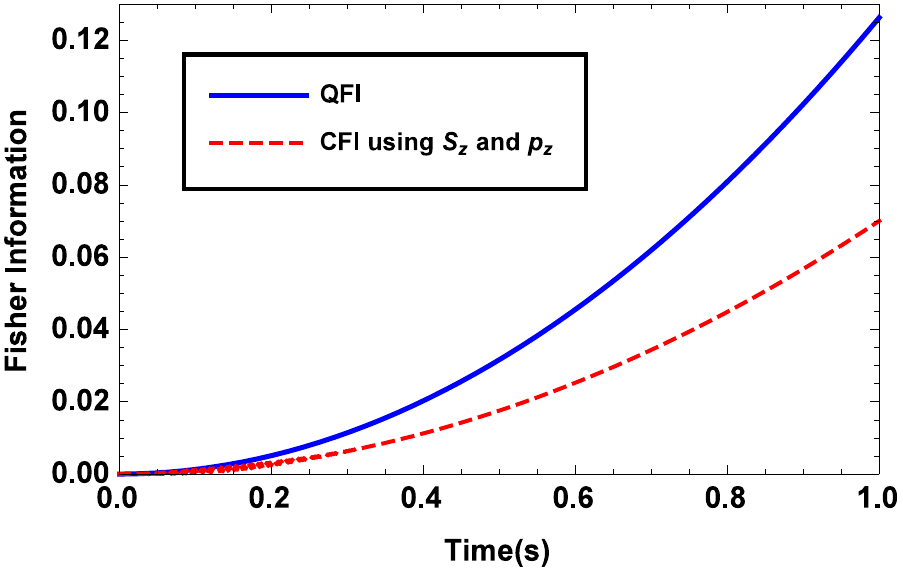}
\caption{Fisher Information with kinetic energy while measuring $S_{z}$ and
$p_{z}$.}
\label{Fig7}
\end{figure}

Then we can get the integral formalism of CFI for population-difference
measurement. Nevertheless, in this case, the CFI is extremely small
compared to the QFI, which means the population-difference measurement
is not a good candidate for estimating $\Omega$. However, since the
QFI, as the maximum of CFI, is large, we can believe some measurement
should facilitate the ascent of CFI. If we carry out a combined measurement (CM) that
resolves the internal states and the momentum distribution $\hat{\Lambda}=(\hat{S}_{z},\hat{p})$, the CFI is
\begin{equation}
I_{CM}=\sum_{s=a,b}\int\mathrm{d}p\frac{[\partial_{\Omega}P_{s}(p)]^{2}}{P_{s}(p)},
\end{equation}
where
\begin{equation}
P_{s}(p)=|\langle\psi_{0}|p\rangle|^{2}(\cos^{2}c'+n_{z}'^{2}\sin^{2}c').
\end{equation}
As a result of this kind of measurement, we continue to maintain $\ k_{0}=10^{6}\ \mathrm{m^{-1}}$, $\Delta=10^{3}\ \mathrm{s^{-1}}$, and we show the CFI becomes larger in figure \ref{Fig7}, meaning we will gain more information about the parameter.

Notably, this is the region where the kinetic energy term cannot be neglected, so it is just an illusion that using population-difference measurement alone can saturate the maximal QFI according to the last section. As a result, when we take the kinetic energy into account, we need to choose different measurement schemes to reach or just approach the maximal QFI. Our result indicates that the neglect of the kinetic energy term will make people carry out incorrect measurements, which they thought has attained the maximal measuring accuracy of Rabi frequency, whereas they are not in some regimes.
\section{Conclusion}\label{sec:6}
In summary, we have theoretically investigated the effect of kinetic
energy on the measuring precision of the Rabi frequency. Through transforming
the original Hamiltonian into a $su$(2)-type Hamiltonian, we can obtain
the analytical expressions of the QFI and CFI indicating the frequency uncertainty.

We now list our main results in the following. Firstly, only when
the absolute value of the detuning or the wave vector is small enough can we safely omit
the kinetic energy term, which we can conclude from both the fidelity and
the QFI. Meanwhile, reducing the two parameters can enhance the QFI in
cases with or without the kinetic energy in Hamiltonian. Secondly,
the inappropriate omission of the kinetic energy term will lead to
overestimating the measuring accuracy of Rabi frequency when $\Delta>0$ or inducing underestimating the precision when $\Delta$ is smaller than some negative value at the given wave vector. Thirdly, a particular initial state's variance in position space can attain a maximum in QFI while taking the kinetic energy term into account. Lastly, although we found that the population-difference measurement is not adequate for CFI with the kinetic energy term, inclusion of the momentum measurement will help a lot.

The Fisher information in different regions of parameters can be deduced from our expressions, and they can contribute to the precise measurement of the Rabi frequency in realistic situations. For example, to improve the measuring accuracy, experimentalists can use our method to calculate the QFI in their interested regime and determine whether the kinetic energy term can be safely omitted. If the parameters do not satisfy the valid neglect, observers need to take care of the schemes of measurement since they may not optimize the precision. Our result may exhibit its value especially in cold atom and atom interferometry \cite{2,42}.

\ack
Xiaoguang Wang proposed the idea and Xingyu Zhang did the calculations
and wrote this article. This work was supported by the National Key
Research and Development Program of China (Grants No.2017YFA0304202
and No.2017YFA0205700), the NSFC (Grants No.11875231 and No.11935012),
and the Fundamental Research Funds for the Central Universities through
Grant No.2018FZA3005.The authors declare that they have no conflict
of interests.

\appendix
\section{Detailed calculation of the quantum Fisher information}\label{appendixA}
\def\thesubsection{\Arabic\alph{section}}

We use the crucial result of the article \cite{14}. For the $su$(2)-type
Hamiltonian $\hat{H}=\vec{r}(\phi)\cdot\vec{J}$, we focus on a unitary
parametrization $\hat{U}=\mathrm{e}^{-\mathrm{i}\hat{H}(\phi)t}$ (here we set $\hbar=1$
and we can recover it after the whole calculation using dimension
analysis). Then we can define a Hermitian operator $\mathcal{H}=\mathrm{i}(\partial_{\phi}\hat{U}^{\dagger})\hat{U}$
and a velocity vector $\vec{v}=\frac{\mathrm{d}\vec{r}}{\mathrm{d}\theta}$.
For an initial pure state $|\psi\rangle$, we have the QFI for the final state: $F=4(\langle\mathcal{H}^{2}\rangle-\langle\mathcal{H}\rangle^{2})$,
where the
\begin{eqnarray}
\mathcal{H}
&=&
\frac{(\vec{r}\cdot\vec{v})(\sin(|\vec{r}|t)-|\vec{r}|t)}{|\vec{r}|^{3}}\vec{r}\cdot\vec{J}-\frac{\sin(|\vec{r}|t)}{|\vec{r}|}\vec{v}\cdot\vec{J}\nonumber\\
&&+\frac{1-\cos(|\vec{r}|t)}{|\vec{r}|^{2}}(\vec{r}\times\vec{v})\cdot\vec{J}\nonumber\\
&=&
\vec{R}\cdot\vec{\sigma}.
\end{eqnarray}

Case 1: the QFI with the kinetic energy
\begin{eqnarray}
\mathcal{\mathcal{H}}_{1} & =\mathrm{i}(\partial_{\Omega}\hat{U}_{1}^{\dagger})\hat{U}_{1}=\mathrm{i}\hat{W}(\partial_{\Omega}\mathrm{e}^{\mathrm{i}t\hat{H}_{e1}})
\mathrm{e}^{-\mathrm{i}t\hat{H}_{e1}}\hat{W}^{\dagger}\nonumber \\
 & =\mathrm{i}\hat{W}(\partial_{\Omega}\mathrm{e}^{\mathrm{i}t(\vec{r}\cdot\vec{J})})\mathrm{e}^{-\mathrm{i}t(\vec{r}\cdot\vec{J})}\hat{W}^{\dagger}\nonumber \\
 & =\hat{W}\mathcal{H}^{'}\hat{W}^{\dagger},
\end{eqnarray}
where $\vec{r}=(\Omega,0,\frac{k_{0}}{m}\hat{p}+\Delta)$, $\vec{v}=(1,0,0)$.

So $\mathcal{H}^{'}=\vec{R}\cdot\vec{\sigma}$,
\begin{eqnarray}
\vec{R}=&\biggl(\frac{\Omega^{2}(\sin(|\vec{r}|t)-|\vec{r}|t)}{2|\vec{r}|^{3}}-\frac{\sin(|\vec{r}|t)}{2|\vec{r}|},\nonumber\\
&\frac{1-\cos(|\vec{r}|t)}{2|\vec{r}|^{2}}(\frac{k_{0}}{m}\hat{p}+\Delta),
\frac{\Omega(\sin(|\vec{r}|t)-|\vec{r}|t)}{2|\vec{r}|^{3}}(\frac{k_{0}}{m}\hat{p}+\Delta)\biggr).
\end{eqnarray}

We use the following initial state:
\begin{eqnarray}
\langle p|\psi_{0}\rangle & =\int_{-\infty}^{\infty}\langle p|z\rangle\langle z|\psi_{0}\rangle=\frac{\sqrt{\sigma}}{\sqrt[4]{\pi}}\mathrm{e}^{-\frac{p^{2}\sigma^{2}}{2\hbar^{2}}}.
\end{eqnarray}
Then,
\begin{eqnarray}
\hat{W}^{\dagger}|a,\psi_{0}\rangle & =\int_{-\infty}^{\infty}\mathrm{e}^{\mathrm{i}\frac{k_{0}\hat{z}}{2}}|p\rangle\mathrm{d}p\langle p|\psi_{0}\rangle|a\rangle\nonumber \\
 & =\int_{-\infty}^{\infty}\mathrm{d}p\frac{\sqrt{\sigma}}{\sqrt[4]{\pi}}\mathrm{e}^{-\frac{p^{2}\sigma^{2}}{2\hbar^{2}}}\mathrm{e}^{\mathrm{i}\frac{k_{0}\hat{z}}{2}}|p\rangle|a\rangle,
\end{eqnarray}
we can simplify the above as
\begin{eqnarray}
\hat{W}^{\dagger}|a,\psi_{0}\rangle & =\int_{-\infty}^{\infty}\mathrm{d}p\frac{\sqrt{\sigma}}{\sqrt[4]{\pi}}\mathrm{e}^{-\frac{p^{2}\sigma^{2}}{2\hbar^{2}}}|p+\frac{\hbar k_{0}}{2}\rangle|a\rangle.
\end{eqnarray}

Finally we obtain
\begin{eqnarray}
\langle\mathcal{\mathcal{H}}_{1}\rangle & =-\frac{\sigma}{\sqrt{\pi}}\int_{-\infty}^{\infty}\mathrm{d}p\mathrm{e}^{-\frac{p^{2}\sigma^{2}}{\hbar^{2}}}\langle p+\frac{\hbar k_{0}}{2}|\hat{R}_{z}|p+\frac{\hbar k_{0}}{2}\rangle,
\end{eqnarray}
and

\begin{eqnarray}
\langle\mathcal{H}_{1}^{2}\rangle & =\frac{\sigma}{\sqrt{\pi}}\int_{-\infty}^{\infty}\mathrm{d}p\mathrm{e}^{-\frac{p^{2}\sigma^{2}}{\hbar^{2}}}\langle p+\frac{\hbar k_{0}}{2}|\hat{R}^{2}|p+\frac{\hbar k_{0}}{2}\rangle.
\end{eqnarray}

\begin{eqnarray}
F= & 4\Biggl[\frac{\sigma}{\hbar\sqrt{\pi}}\int_{-\infty}^{\infty}\mathrm{d}p\ \mathrm{e}^{-\frac{p^{2}\sigma^{2}}{\hbar^{2}}}\Biggl\{\left(\frac{\Omega^{2}\left(\sin\left(\omega t\right)-\omega t\right)}{2\omega^{3}}-\frac{\sin\left(\omega t\right)}{2\omega}\right)^{2}\nonumber \\
 &+ \left(\frac{\sqrt{\omega^2-\Omega^2}\left(1-\cos\left(\omega t\right)\right)}{2\omega^{2}}\right)^{2}+\left(\frac{\Omega\sqrt{\omega^2-\Omega^2}(\sin\left(\omega t\right)-\omega t)}{2\omega^{3}}\right)^{2}\Biggl\}\Biggl]\nonumber \\
 &- 4\Biggl[\frac{\sigma}{\hbar\sqrt{\pi}}\int_{-\infty}^{\infty}\mathrm{d}p\ \mathrm{e}^{-\frac{p^{2}\sigma^{2}}{\hbar^{2}}}\Biggl(\frac{\Omega\sqrt{\omega^2-\Omega^2}(\sin\left(\omega t\right)-\omega t)}{2\omega^{3}}\Biggr)\Biggr]^{2}.
\end{eqnarray}
where
\begin{equation}
\omega=\sqrt{\left(\Delta+\frac{k_{0}\left(\frac{k_{0}\hbar}{2}+p\right)}{m}\right)^{2}+\Omega^{2}}.
\end{equation}

Case 2: the QFI without the kinetic energy

\begin{eqnarray}
\mathcal{H}_{2} & =\mathrm{i}(\partial_{\Omega}\hat{U}_{2}^{\dagger})U_{2}=\mathrm{i}\hat{W}(\partial_{\Omega}\mathrm{e}^{\mathrm{i}t\hat{H}_{e2}})\mathrm{e}^{-\mathrm{i}t\hat{H}_{e2}}\hat{W}^{\dagger}\nonumber \\
 & =\mathrm{i}\hat{W}(\partial_{\Omega}\mathrm{e}^{\mathrm{i}t(\vec{r}\cdot\vec{J})})\mathrm{e}^{-\mathrm{i}t(\vec{r}\cdot\vec{J})}\hat{W}^{\dagger}\nonumber \\
 & =\hat{W}\mathcal{H}^{'}\hat{W}^{\dagger},
\end{eqnarray}
where $\vec{J}=\frac{\vec{\sigma}}{2}$, $\vec{r}=(\Omega,0,\Delta)$,
$\vec{v}=\frac{\mathrm{d}\vec{r}}{\mathrm{d}\Omega}=(1,0,0)$.

\begin{eqnarray}
\mathcal{H}^{'}&=&\frac{\Omega(\sin(|\vec{r}|t)-|\vec{r}|t)}{|\vec{r}|^{3}}(\frac{\Omega}{2}\sigma_{x}+\frac{\Delta}{2}\sigma_{z})\nonumber \\
 &&-\frac{\sin(|\vec{r}|t)}{|\vec{r}|}\frac{\sigma_{x}}{2}+\frac{1-\cos(|\vec{r}|t)}{|\vec{r}|^{2}}\frac{\Delta}{2}\sigma_{y}\nonumber \\
 &=&\vec{R}\cdot\vec{\sigma},
\end{eqnarray}
where

\begin{eqnarray}
\vec{R}= & \biggl(\frac{\Omega^{2}(\sin(|\vec{r}|t)-|\vec{r}|t)}{2|\vec{r}|^{3}}-\frac{\sin(|\vec{r}|t)}{2|\vec{r}|},\nonumber \\
 & \frac{1-\cos(|\vec{r}|t)}{2|\vec{r}|^{2}}\Delta,\frac{\Omega(\sin(|\vec{r}|t)-|\vec{r}|t)}{2|\vec{r}|^{3}}\Delta\biggr).
\end{eqnarray}

Since the Fisher Information $F=4(\langle\mathcal{H}_{2}^{2}\mathcal{\rangle}-\langle\mathcal{H}_{2}\rangle^{2})$,
the initial state is $|a,\psi_{0}\rangle$, where $\langle z|\psi_{0}\rangle=\frac{1}{\sqrt[4]{\pi\sigma^{2}}}\exp(-\frac{z^{2}}{2\sigma^{2}})$.
Then due to $\hat{W}^{\dagger}|a,\psi_{0}\rangle=|a\rangle \mathrm{e}^{\mathrm{i}\frac{k_{0}\hat{z}}{2}}|\psi_{0}\rangle$,
the z-space doesn't contribute to the final Fisher information.

\begin{eqnarray}
F=\frac{\left(\Omega^{2}\omega't+\Delta^{2}\sin\left(\omega't\right)\right)^{2}+\Delta^{2}\omega'^{2}\left(\cos\left(\omega't\right)-1\right)^{2}}{\omega'^{6}},
\end{eqnarray}
where
\begin{equation}
\omega'=\sqrt{\Delta^{2}+\Omega^{2}}.
\end{equation}

\section{Detailed Derivation of the fidelity}\label{appendixB}
\def\thesubsection{\Arabic\alph{section}}
In this appendix, we show how the fidelity (\ref{eq:fidelity}) is calculated in detail.
\begin{eqnarray}
\mathcal{F} & =|\langle\psi_{0}|\langle a|\hat{W}\mathrm{e}^{\mathrm{i}\frac{t}{\hbar}\hat{H}_{e1}}\mathrm{e}^{-\mathrm{i}\frac{t}{\hbar}\hat{H}_{e2}}\hat{W}^{\dagger}|a\rangle|\psi_{0}\rangle|^{2}\nonumber\\
 & =|\langle\psi_{0}|\langle a|\mathrm{e}^{-\mathrm{i}\frac{k_{0}\hat{z}}{2}}\mathrm{e}^{\mathrm{i}\frac{t}{\hbar}\hat{H}_{e1}}\mathrm{e}^{-\mathrm{i}\frac{t}{\hbar}\hat{H}_{e2}}\mathrm{e}^{\mathrm{i}\frac{k_{0}\hat{z}}{2}}|a\rangle|\psi_{0}\rangle|^{2}\nonumber\\
 & =|\langle\psi_{0}|\mathrm{e}^{-\mathrm{i}\frac{k_{0}\hat{z}}{2}}\mathrm{e}^{\mathrm{i}\frac{t}{\hbar}(\frac{\hat{p}^{2}}{2m}+\frac{\hbar^{2}k_{0}^{2}}{8m})}\langle a|\mathrm{e}^{\mathrm{i}\frac{t}{\hbar}[\frac{\hbar\Omega}{2}\sigma_{x}+(\frac{\hbar k_{0}\hat{p}}{2m}+\frac{\hbar\Delta}{2})\sigma_{z}]}\mathrm{e}^{-\mathrm{i}\frac{t}{\hbar}[\frac{\hbar\Omega}{2}\sigma_{x}+\frac{\hbar\Delta}{2}\sigma_{z}]}|a\rangle \mathrm{e}^{\mathrm{i}\frac{k_{0}\hat{z}}{2}}|\psi_{0}\rangle|^{2}\nonumber\\
 &~
\end{eqnarray}

We set $c=\frac{t}{\hbar}\sqrt{(\frac{\hbar k_{0}\hat{p}}{2m}+\frac{\hbar\Delta}{2})^{2}+(\frac{\hbar\Omega}{2})^{2}}$,
$\vec{m}=(\frac{\frac{\hbar\Omega}{2}}{\sqrt{(\frac{\hbar k_{0}\hat{p}}{2m}+\frac{\hbar\Delta}{2})^{2}+(\frac{\hbar\Omega}{2})^{2}}},0,\frac{\frac{\hbar k_{0}\hat{p}}{2m}+\frac{\hbar\Delta}{2}}{\sqrt{(\frac{\hbar k_{0}\hat{p}}{2m}+\frac{\hbar\Delta}{2})^{2}+(\frac{\hbar\Omega}{2})^{2}}})$, and using the exponential expand of the pauli matrix, we can get the expression:
\begin{equation}
\mathrm{e}^{\mathrm{i}\frac{t}{\hbar}[\frac{\hbar\Omega}{2}\sigma_{x}+(\frac{\hbar k_{0}\hat{p}}{2m}+\frac{\hbar\Delta}{2})\sigma_{z}]}=\cos c+\mathrm{i}(m_{1}\sigma_{x}+m_{3}\sigma_{z})\sin c.
\end{equation}
Similarly, we set $d=-\frac{t}{\hbar}\sqrt{(\frac{\hbar\Delta}{2})^{2}+(\frac{\hbar\Omega}{2})^{2}}$,
$\vec{n}=(\frac{\frac{\hbar\Omega}{2}}{\sqrt{(\frac{\hbar\Delta}{2})^{2}+(\frac{\hbar\Omega}{2})^{2}}},0,\frac{\frac{\hbar\Delta}{2}}{\sqrt{(\frac{\hbar\Delta}{2})^{2}+(\frac{\hbar\Omega}{2})^{2}}})$
and substitute these into the other complex exponential expand:
\begin{equation}
\mathrm{e}^{-\mathrm{i}\frac{t}{\hbar}[\frac{\hbar\Omega}{2}\sigma_{x}+\frac{\hbar\Delta}{2}\sigma_{z}]}=\cos d+\mathrm{i}(n_{1}\sigma_{x}+n_{3}\sigma_{z})\sin d.
\end{equation}

Then the inner product of the internal states can be calculated as
\begin{eqnarray}
 & \langle a|\mathrm{e}^{\mathrm{i}\frac{t}{\hbar}[\frac{\hbar\Omega}{2}\sigma_{x}+(\frac{\hbar k_{0}\hat{p}}{2m}+\frac{\hbar\Delta}{2})\sigma_{z}]}\mathrm{e}^{-\mathrm{i}\frac{t}{\hbar}[\frac{\hbar\Omega}{2}\sigma_{x}+\frac{\hbar\Delta}{2}\sigma_{z}]}|a\rangle\nonumber\\
= & \langle a|[\cos c+\mathrm{i}(m_{1}\sigma_{x}+m_{3}\sigma_{z})\sin c]|a\rangle\langle a|[\cos d+\mathrm{i}(n_{1}\sigma_{x}+n_{3}\sigma_{z})\sin d]|a\rangle\nonumber\\
 &+ \langle a|[\cos c+\mathrm{i}(m_{1}\sigma_{x}+m_{3}\sigma_{z})\sin c]|b\rangle\langle b|[\cos d+\mathrm{i}(n_{1}\sigma_{x}+n_{3}\sigma_{z})\sin d]|a\rangle\nonumber\\
= & (\cos c-\mathrm{i}m_{3}\sin c)(\cos d-\mathrm{i}n_{3}\sin d)+\mathrm{i}m_{1}\sin c\ \mathrm{i}n_{1}\sin d\nonumber\\
= & \cos c\cos d-\mathrm{i}m_{3}\sin c\cos d-\mathrm{i}n_{3}\sin d\cos c\nonumber\\
&-m_{3}\sin c\ n_{3}\sin d-m_{1}\sin c\ n_{1}\sin d\nonumber\\
= & \cos c\cos d-\mathrm{i}m_{3}\sin c\cos d\nonumber\\
&-\mathrm{i}n_{3}\sin d\cos c-(m_{3}n_{3}+m_{1}n_{1})\sin c\sin d.
\end{eqnarray}

Now we see that the overlap between the two evolved final state is
\begin{eqnarray}
 &\langle\psi_{0}|\mathrm{e}^{-\mathrm{i}\frac{k_{0}\hat{z}}{2}}\mathrm{e}^{\mathrm{i}\frac{t}{\hbar}(\frac{\hat{p}^{2}}{2m}+\frac{\hbar^{2}k_{0}^{2}}{8m})}\langle a|\mathrm{e}^{\mathrm{i}\frac{t}{\hbar}[\frac{\hbar\Omega}{2}\sigma_{x}+(\frac{\hbar k_{0}\hat{p}}{2m}+\frac{\hbar\Delta}{2})\sigma_{z}]}\mathrm{e}^{-\mathrm{i}\frac{t}{\hbar}[\frac{\hbar\Omega}{2}\sigma_{x}+\frac{\hbar\Delta}{2}\sigma_{z}]}|a\rangle \mathrm{e}^{\mathrm{i}\frac{k_{0}\hat{z}}{2}}|\psi_{0}\rangle\nonumber\\
=&\int\mathrm{d}p|\langle\psi_{0}|p\rangle|^{2}[\mathrm{e}^{\mathrm{i}\frac{t}{\hbar}(\frac{\hat{p}^{2}}{2m}+\frac{\hbar^{2}k_{0}^{2}}{8m})}(\cos c\cos d-\mathrm{i}m_{3}\sin c\cos d\nonumber\\
&-\mathrm{i}n_{3}\sin d\cos c-(m_{3}n_{3}+m_{1}n_{1})\sin c\sin d)],
\end{eqnarray}
where $\hat{p}$ is replaced by $p+\frac{\hbar k_{0}}{2}$ ($c$ also includes $\hat{p}$). Then the modulus square, fidelity $\mathcal{F}$, can be numerically integrated.

\end{document}